# Social Computing Based Analysis on Monogamous Marriage Puzzle of Human

CAI, Ning[1, 2]    DIAO, Chen[1, 2]    YAN, Bo-Han[3]    LIU, Jin-Hu[4]

[1] School of Electrical Engineering, Northwest University for Nationalities, Lanzhou, China
[2] Key Laboratory of National Language Intelligent Processing, Gansu Province, China
[3] School of Ethnology and Sociology, Northwest University for Nationalities, Lanzhou, China
[4] People's Hospital of Linxia Prefecture, Linxia, China

**Abstract:** Most of the mammal species hold polygynous mating systems. The majority of the marriage systems of mankind were also polygynous over civilized history, however, socially imposed monogamy gradually prevails throughout the world. This is difficult to understand because those mostly influential in society are themselves benefitted from polygyny. Actually, the puzzle of monogamous marriage could be explained by a simple mechanism, which lies in the sexual selection dynamics of civilized human societies, driven by wealth redistribution. The discussions in this paper are mainly based on the approach of social computing, with a combination of both experimental and analytical analysis.

**Key Words:** Marriage Systems; Monogamy; Polygyny; Social Computing

## 1. Introduction

It is estimated that up to 90% mammal species have polygynous mating systems [1], including human, in which one male lives and mates with multiple females, whereas each female only mates with a single male. This is evidently advantageous for facilitating evolution by effective sexual selection.

In comparison with most of other mammals, where generally any species possesses a constant unitary mating system, extensive complexity is manifested in human. Nearly all typical modes exist in anthropological records, e.g. monogamy, polygyny, polyandry, and promiscuity. Nonetheless, polygyny is still dominant throughout civilizations [2], which also conforms to certain biological signs such as the moderate sexual dimorphism in body size [3].

---

[1] Corresponding author: Cai, Ning (caining91@tsinghua.org.cn)

Despite the above facts, during the recent centuries, socially imposed monogamy which is regulated both by laws and ethics gradually prevails all over the world. The cause of such a phenomenon seems somewhat abstruse, mainly because those who are most influential in establishing laws and shaping norms are exactly the same stratum being mostly benefitted from polygyny [4-6]. It is difficult to understand why these people voluntarily abandon the privilege in holding more than one wife. Thus, it is called a puzzle [4].

There exist several hypotheses in the literature for the drive of the transition from polygyny to socially imposed monogamy, e.g. male power dynamics, technological impacts, cultural group selection, pathogen stress, and inclusive fitness [4, 7-11]. Recently, Bauch and McElreath [12] proposed a novel hypothesis that monogamy surpasses polygyny primarily in reducing the negative effects from sexually transmitted bacterial infections.

The hypothesis of Bauch and McElreath is theoretically reasonable and enlightening, however, it may not always be consistent with the reality. First, although agriculture has developed for thousands of years, socially imposed monogamy reigned over the world only during the very recent centuries. For instance, polygyny was legal before 1880 in Japan, 1953 in China, 1955 in India, and 1963 in Nepal [13]. Records can hardly be found to show any sign of correlation between depopulation and sexually transmitted infections in these Asian countries. Secondly, mating system is different from pair bond marriage system. Even if the monogamous marriage system is validly implemented, sexually transmitted infections can still spread via a hidden dissimilar mating network, e.g. prostitution.

In this article, we propose a new hypothesis, which is a simple mechanism to explain the monogamous marriage puzzle. We speculate that the de-facto monogamy may not be the result of initiative rational choice of any people; instead, it should be naturally yielded from the sexual selection dynamics of mankind in civilized societies. The key of the mechanism lies in the redistribution of wealth.

The discussions are mainly based on the approach of social computing, with a combination of both empirical and analytical analysis, in which the empirical analysis here is rooted within the theoretical framework of parallel systems [14-15]. We study the laws of social phenomena by building and observing the behaviors of simulation systems. The objective of simulation systems is not for comprehensively and quantitatively mimicking the real society, instead, it could be very conducive to drawing conclusions about certain issues qualitatively, especially those negative conclusions asserting that something should never occur.

The rest of this paper is organized as follows. Section 2 elaborates the mechanism of the monogamous tendency based on an agent-based simulation model. Section 3 particularly discusses the effect of marriage system to the intensity of the overall wealth gap in a society. The impact of sexually transmitted diseases to the population is analyzed under different marriage systems in Section 4. Finally, Section 5 presents the concluding remarks.

## 2. Simple Mechanism to Explain the Puzzle

As a society shifted into patriarchal, it must adopt either polygamous or monogamous marriage system; meanwhile develop a set of rigorous norms to compel women to keep their chastity. This is because a father needs to guarantee the authenticity of his parenthood, in order that his property can be inherited by the genuine children of his own.

The spouse selection in civilized society is generally economy driven. Suppose that at an initial stage, the marriage system is polygamous. Polygyny is dependent on the intensity of overall wealth gap in a society. Intensive polygyny should be rooted on an intensive status of wealth gap. A prerequisite for someone to keep more wives should be that he is richer than average. However, having more wives means having more children, and accordingly, the inheritance of his property would be diluted since it has to be divided into more number of parcels. Such a mechanism could naturally suppress the wealth gap and the difference of wife numbers throughout a society, correspondingly.

Polygyny can effectively limit the accumulation of property through generations and suppress the overall status of wealth gap. This is a spontaneous mechanism of balance, with dynamic negative feedback. A man who keeps multiple wives enjoys the welfare of being advantageous in spreading his genes, but he must also bear the punishment of diluting property by inheritance. In this way, the advantage of his descendants for competetion in sexual selection is weakened, as compared with himself.

The mechanism restraining the intensity of polygyny is empirically testified by a very simple agent-based model. One will see that the variance of wife number keeps approximating zero, indicating almost de-facto monogamy.

The model is discrete-timed, with each iteration representing a generation of people. The procedure of the model can be divided into several sections.

The first section is wealth distribution. The initial wealth distribution among men is Gaussian. Assume that initially, the overall wealth gap is relatively higher. This might be due to a war just occurred, or some other event that could arouse redistribution of social wealth. The serious wealth gap is expressed by randomly selecting 10% men to hold additional wealth, with the mathematical expectation many times greater than ordinary.

The second section is marriage, which is the most important section. The number of wives of a man is determined by the wealth he holds. Such a relation between the quantity of wives and the amount of wealth relative to others is depicted by a function in our model, which has several principles:

1) The function is increasing.

2) The function will tend to some quasi-saturation if the amount of wealth is sufficiently great. The reason is due to both the bounded demand and the limited resource.

3) The slope of the function is not only less than 1 in general, but also decreasing. This analogously accords with the polygyny threshold model [16] observed in animals.

4) The value of the function is 1 as the amount of wealth equals the average.

According to the above principles, we set the relation function as the following form.

$$y = \lambda \tanh(x/\mu) - \lambda \tanh(1/\mu) + 1 \qquad (1)$$

where $\tanh(\bullet)$ is the hyperbolic tangent function; $x \in R$ is the ratio of the amount of wealth to the average; $y \in R^+$ is the mathematical expectation of wife quantity; and $\lambda, \mu \in R^+$ are the parameters shaping the curve.

The program randomly chooses an unmarried man and assigns him a number of wives from the set of unmarried women in the same generation. For this purpose, firstly function (1) is computed. Next, the idiosyncratic effect due to other factors except wealth is reflected by additionally multiplying a noise, which is a Gaussian random number with the mathematical expectation being 1. Finally, the rounded result is the quantity of his wives. After this, the program randomly chooses another unmarried man to assign him wives, similarly following the rules above. Such a process is repeated until all the unmarried women are assigned out.

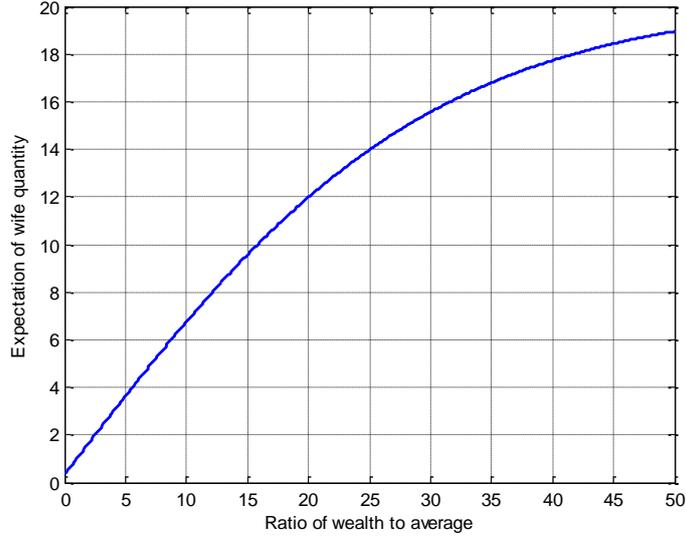

Fig. 1. A sample of relation between ratio of wealth to average and mathematical expectation of wife quantity

The third section is reproduction. It is rational to set the average fertility of women to 3 or 4, representing the quantity of children that could grow up. There are different cases in different families. Thus, for each woman, the actual quantity should be yielded by multiplying an additional noise, which is a Gaussian random number with the mathematical expectation being 1.

The fourth section is inheritance. Since it is patriarchal society, the wealth of a father would be evenly divided and inherited by his sons. This is consistent with the very situation in ancient China [17]. In addition, each of the sons will have a career, and his life savings by his own should be added, which is also modeled as a Gaussian random number.

For the beginning generation, there are equal numbers of men and women. Then the new generations are iteratively reproduced, following the same procedure described above.

We conducted experiments based on the model and observed the results. The variance of the wife quantities derived via the following formula

$$\frac{\sum_{i=1}^{N_m}(q_i-1)^2}{N_m-1} \tag{2}$$

is taken as the index indicating the intensity of polygyny in the overall society, since the mean of wife quantities is always 1. Note that in (2), $N_m$ denotes the population of men and $q_i$ denotes the wife quantity of the $i$th man. The greater this variance, the higher level of polygyny, or in other word, the lower level of monogamy happens.

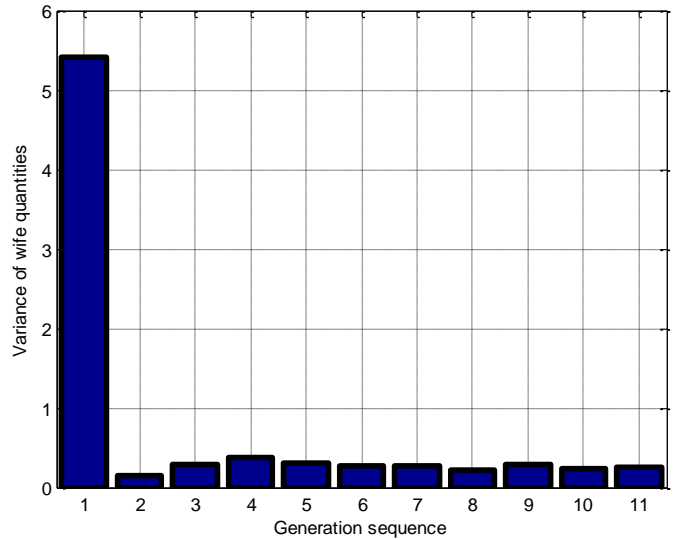

Fig. 2. Variance of wife quantities

According to the experimental results, a primary conclusion is that in the steady state after the first few generations, the intensity of polygyny keeps very low, with the variance being close to zero. An example is illustrated in Fig. 2, which also manifests an interesting phenomenon in our experiments that ultra high level of monogamy usually occurs in the generation being subsequent to any generation bearing very serious wealth gap. For a more intuitive perception of such a variance, four typical fragments of wife quantities in different generations are extracted from experimental results and listed in Tab. 1, each with 30 successive samples.

Tab. 1. Fragments of wife quantities of four generations

| Generation sequence | Fragment of wife quantities |
|---|---|
| 1 | 14, 0, 1, 1, 0, 0, 1, 4, 1, 1, 0, 1, 0, 0, 0, 1, 1, 0, 0, 1, 0, 0, 1, 1, 5, 0, 0, 0, 0, 0 |
| 2 | 1, 1, 1, 1, 1, 2, 1, 0, 1, 1, 1, 1, 1, 1, 1, 0, 1, 0, 0, 1, 1, 1, 1, 1, 1, 1, 1, 1, 2, 1 |
| 3 | 1, 1, 1, 1, 1, 1, 1, 2, 1, 1, 1, 0, 1, 1, 1, 1, 1, 1, 1, 1, 1, 1, 1, 1, 1, 2, 1, 1, 1 |
| 4 | 1, 1, 1, 1, 0, 1, 1, 1, 1, 1, 0, 1, 1, 0, 1, 1, 1, 1, 1, 1, 1, 2, 1, 1, 1, 2, 1, 1, 1, 0 |

## 3. Effect to Wealth Gap

The fact that polygyny contributes to supress the overall wealth gap in society can also be clearly verified by experimental observations.

Fig. 3 illustrates a comparison of overall wealth gaps between polygynous and monogamous societies under the current model. Here the intensity of wealth gap is measured by the ratio of standard deviation to average wealth, being expressed as

$$\left.\sqrt{\frac{\sum_{i=1}^{N_m}(\zeta_i - \zeta_a)^2}{N_m - 1}}\right/\zeta_a \qquad (3)$$

where $\zeta_i$ is the amount of wealth held by the *i*th man, and $\zeta_a$ is the average wealth over the society.

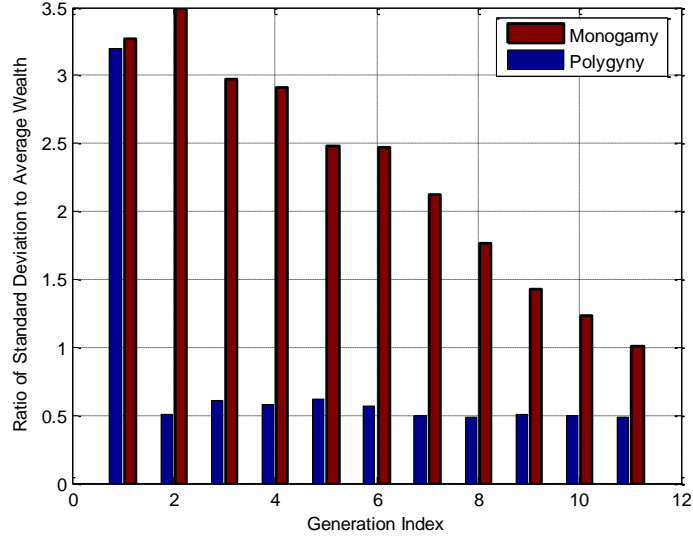

Fig. 3. Comparison of overall wealth gaps between polygynous and monogamous societies

One can see that the intensity of wealth gap generally keeps very low and relatively stationary in a polygynous society; whereas in contrast, it is significantly higher in a society with imposed monogamy. After generations, such a difference declines. This may be attributed to the natural tendency of de-facto monogamy.

## 4. Effect of Sexually Transmitted Diseases

Recently, Bauch and McElreath proposed a novel hypothesis in [12]. Based on an empirical model, they conjecture that as the scale of community kept on growing after the origin of agriculture, the impact of STD on the overall fertility of the polygamists became serious; and thereby the monogamists finally dominated in population.

Actually, a simple analytical analysis can be conducted to help validating and clarifying the hypothesis of Bauch and McElreath.

Suppose that the total population of the *k*th generation is $\xi(k)$, with the women/men population being $\xi(k)/2$; the baseline birth rate of a woman is denoted by $\alpha$, which is the number of children in her lifetime and can be affected by many factors expressed as multiplicative coefficients; the probability of a married woman to be infected by STD from some source outside of family is $\beta$, with the corresponding probability of a man being $\tilde{\beta}$; the probability of sterility for an infected woman is $\gamma$; and the average count of wives in a family is $q$.

STD restrains the increase of population. It can be analytically explained whether or not this effect is intensified by polygyny.

For simplicity, assume that a man and all his wives will eventually get infected if anyone of them is initially infected by STD and becomes infectious.

Consider the factor attributed to an initially infected husband. An expected birth rate of any woman under this factor is

$$\begin{aligned}&\alpha(1-\tilde{\beta})+\alpha(1-\gamma)\tilde{\beta}\\&=\alpha(1-\tilde{\beta}\gamma)\end{aligned} \quad (4)$$

where $\alpha$ is the baseline birth rate excluding the effect of the current factor. This is independent of the count of wives.

Now consider the factor attributed to an initially infected wife. The expected birth rate of any woman under this factor becomes

$$\begin{aligned}&\alpha(1-\beta)^q+\alpha(1-\gamma)[1-(1-\beta)^q]\\&=\alpha\{1+[(1-\beta)^q-1]\gamma\}\end{aligned} \quad (5)$$

where $(1-\beta)^q$ is the probability that none of the wives get infected. The value of (5) is negatively correlated to $q$. Suppose that $\beta=0.04$ and $\gamma=0.2$. The relation between the coefficient $1+[(1-\beta)^q-1]\gamma$ and $q$ is illustrated in Fig. 4.

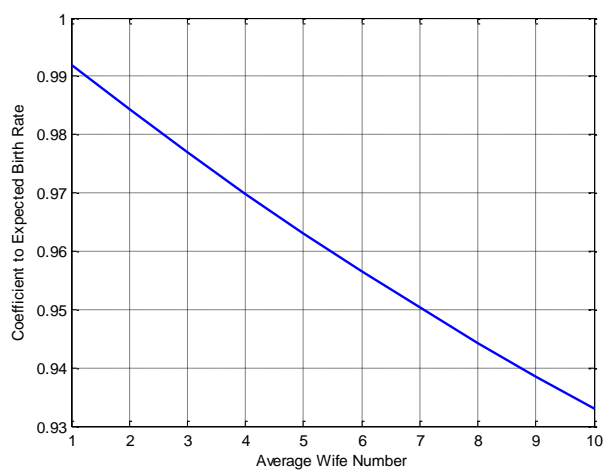

Fig. 4. Relation between average wife number $q$ and coefficient to expected birth rate $1+[(1-\beta)^q-1]\gamma$. $\beta=0.04$ and $\gamma=0.2$.

Under the factor of an initially infected wife, the population dynamics are depicted by the following equations:

$$\xi(k+1) = \frac{\alpha}{2}\{1+[(1-\beta)^q-1]\gamma\}\xi(k) \quad (k=1,2,3...) \tag{6}$$

and

$$\xi(k) = (\frac{\alpha}{2})^{k-1}\{1+[(1-\beta)^q-1]\gamma\}^{k-1}\xi(1) \tag{7}$$

The population is an exponential function of time, with the initial difference being amplified with time. In this way, the population of monogamous community tends to preponderate gradually. Fig. 5 manifests the ratio of monogamous to polygamous population i.e.

$$(\frac{1-\beta\gamma}{1+[(1-\beta)^q-1]\gamma})^{k-1}$$

over generations, where $q = 8$, $\beta = 0.04$, and $\gamma = 0.2$ & $\gamma = 0.4$, respectively.

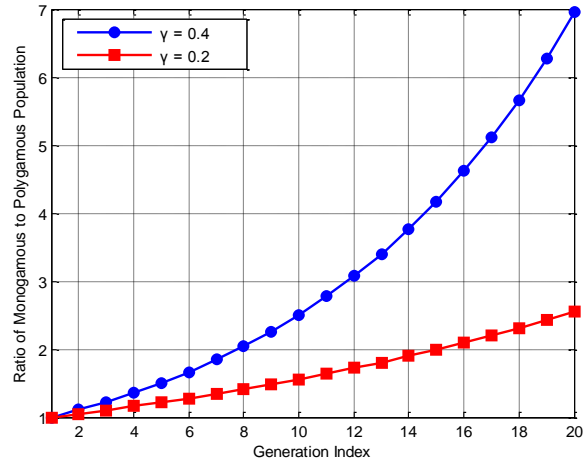

Fig. 5. Ratio of monogamous to polygamous population over generations.

Actually, even the setting of $\gamma = 0.2$ reflects rather high level of STD caused sterility. From Fig. 5, one can see that although due to STD, the population of monogamous community is superior to polygynous, the significance of such an effect may still be comparatively minor in reality.

The most likely cause for a family member to be infected by external STD source is extramarital sex. In any human society, the actual mating system is not always consistent with the marriage system. There exist hidden extramarital relationships such as prostitution. The mechanism analyzed here is compatible with this fact.

# 5. Conclusion

The current article concentrates on the monogamous marriage puzzle of mankind. A simple hypothesis is presented, which is based on an economic perspective. According to the hypothesis, the men who possess multiple wives are confronted with intensive dilution of their wealth via heritage distribution, because they tend to have more children than average people. As a result, the sons are usually less rich than their father and thereby it is difficult for them to keep the same amount of wives. The numerical simulations on an agent-based model clearly manifest consistency with this hypothesis. In the steady state, a very low level of the variance of wife quantity is observable, indicating a de-facto monogamy.

Due to the mechanism summarized above, the de-facto monogamy is a natural tendency for mankind, rather than the rational decision of any group of people. Probably, the overall set of moralities, sentiments, or even laws that have been jointly molding the monogamous pattern might just be the byproducts generated from an adaptation of society to the existing de-facto monogamy.

The simulations also implicates that polygyny effectively suppresses the overall wealth gap in society. This could partially explain why the situation of wealth gap in modern societies is often more serious than history.

It is worth mentioning that the mechanism here is merely hypothesis. In fact, the monogamous marriage puzzle should still be comprehensively attributed to multiple causes, with the mechanism introduced here being a primary factor, so long as it is validated by more evidence.

## Acknowledgments

This work is supported by National Natural Science Foundation (NNSF) of China (Grants 61374054 & 61263002), by Fundamental Research Funds for the Central Universities (Grants 31920160003 & 31920170141), and by Program for Young Talents of State Ethnic Affairs Commission (SEAC) of China (Grant 2013-3-21).

# Appendix (Matlab Code for Review)

```matlab
lamda = 24;
mu = 30;
total_generations = 12;
average_fertility = 3;
men_population_cur_generation = 100;
women_population_cur_generation = 100;
men_population_next_generation = 0;
women_population_next_generation = 0;
total_wealth_cur_generation = 0;
total_wealth_next_generation = 0;
bMonogamy = 1;

%initialization
for i=1:20000
    men_cur_generation(i, 1) = 0; %wealth
    men_cur_generation(i, 2) = 0; %ratio of wealth to average
    men_cur_generation(i, 3) = 0; %quantity of wives
    men_cur_generation(i, 4) = 0; %quantity of sons
    men_next_generation(i) = 0;
end

for i=1:men_population_cur_generation
    men_cur_generation(i, 1) = normrnd(100, 400); %initial wealth
    dice = rand;
    if dice<0.1
        men_cur_generation(i, 1) = men_cur_generation(i, 1)+normrnd(5000, 3000);
    end
    men_cur_generation(i, 3) = 0; %initially unmarried
    men_cur_generation(i, 4) = 0; %initially no children
    total_wealth_cur_generation = total_wealth_cur_generation+men_cur_generation(i, 1);
    men_remain_numbers(i) = i;
end

count = 1;
while count<total_generations

    for i=1:men_population_cur_generation
        men_cur_generation(i, 2) = men_cur_generation(i, 1)/total_wealth_cur_generation*men_population_cur_generation; %ratio of wealth to average
    end

    %marriage match
    women_remain = women_population_cur_generation;
    men_remain = men_population_cur_generation;
    while women_remain > 0
        cur_man = round(rand*men_remain);
        if cur_man == 0
            cur_man = 1;
        end

        %current man is unmarried
        noise = normrnd(1,0.2);
        if bMonogamy == 0
            cur_wives = round((lamda*tanh(men_cur_generation(men_remain_numbers(cur_man), 2)/mu)-lamda*tanh(1/mu)+1)*noise); %expected count of wives for the current man
        else
            cur_wives = 1;
        end
        if cur_wives<=women_remain && cur_wives>0 %marry
            men_cur_generation(men_remain_numbers(cur_man), 3) = cur_wives;
            women_remain = women_remain - cur_wives;
            %remove from list of unmarried
            if men_remain>1
                men_remain = men_remain-1;
            end
```

```
            for i = cur_man:men_remain
                men_remain_numbers(i) = men_remain_numbers(i+1);
            end
        end
    end

    %compute the variance
    variance(count) = 0;
    standard_deviation_wealth(count) = 0;
    average_wealth = total_wealth_cur_generation/men_population_cur_generation;
    for i=1:men_population_cur_generation
        variance(count) = variance(count)+(men_cur_generation(i, 3)-1)^2;
        standard_deviation_wealth(count) =
standard_deviation_wealth(count)+(men_cur_generation(i, 1)-average_wealth)^2;
    end
    variance(count) = variance(count)/(men_population_cur_generation-1);
    standard_deviation_wealth(count) =
sqrt(standard_deviation_wealth(count)/(men_population_cur_generation-1))/average_wealth;

    %birth & inheritage
    cur_young_man = 1;
    men_population_next_generation = 0;
    women_population_next_generation = 0;
    for i=1:men_population_cur_generation
        %birth
         noise = normrnd(1,0.3);
         offspring_num = round(men_cur_generation(i,3)*average_fertility*noise);
        if offspring_num>0
           for j=1:offspring_num
              dice = randint;
              if dice == 0
                 women_population_next_generation = women_population_next_generation+1;
              end
              if dice == 1
                 men_cur_generation(i,4) = men_cur_generation(i,4)+1;
                 men_population_next_generation = men_population_next_generation+1;
              end
           end
        end
        %inheritage
        if men_cur_generation(i,4)>0
           for j=1:men_cur_generation(i,4)
              men_next_generation(cur_young_man) =
men_cur_generation(i,1)/men_cur_generation(i,4)+normrnd(10,8);
              cur_young_man = cur_young_man+1;
           end
        end
    end

    %death
    total_wealth_cur_generation = 0;
    for i=1:men_population_next_generation
       men_cur_generation(i, 1) = men_next_generation(i); %wealth
       men_cur_generation(i, 3) = 0; %initially unmarried
       men_cur_generation(i, 4) = 0; %initially no children
       total_wealth_cur_generation = total_wealth_cur_generation+men_cur_generation(i, 1);
       men_remain_numbers(i) = i;
    end
    men_population_cur_generation = men_population_next_generation;
    women_population_cur_generation = women_population_next_generation;
    count = count+1;
end
```